\documentclass[fdp,a4paper,fleqn,finallayout]{w-art}
\usepackage{times,cite,w-thm}
\theoremstyle{plain}

\theoremstyle{definition}

\newcommand{\be}{\begin{equation}}
\newcommand{\ee}{\end{equation}}
\def\lsi{\raise0.3ex\hbox{$<$\kern-0.75em\raise-1.1ex\hbox{$\sim$}}}
\def\gsi{\raise0.3ex\hbox{$>$\kern-0.75em\raise-1.1ex\hbox{$\sim$}}}
\newcommand{\lsim}{\mathop{\lsi}}
\newcommand{\gsim}{\mathop{\gsi}}
\def\backder{\raise1.4ex\hbox{$\leftarrow$\kern-0.75em\raise-1.4ex\hbox{$\partial$}}}
\def\forder{\raise1.4ex\hbox{$\rightarrow$\kern-0.75em\raise-1.4ex\hbox{$\partial$}}}
\newcommand{\backderi}{\mathop{\backder}}
\newcommand{\forderi}{\mathop{\forder}}
\newcommand{\C}{{\kern+.25em\sf{C}\kern-.45em\sf{{\small{I}}} \kern+.45em\kern-.25em}}
\newcommand{\N}{{\kern+.25em\sf{N}\kern-.8em\sf{I} \kern+.8em\kern-.25em}}
\usepackage[]{graphicx}
\begin{document}
\DOIsuffix{prop.200900021}
\Volume{57}
\Month{3}
\Issue{5-7}
\Year{2009}
\pagespan{505}{513}
\Receiveddate{20 February 2009}
\Accepteddate{27 Febraury 2009}
\Dateposted{31 March 2009}


\title[Non-linear photon dispersion ?]{Could the photon dispersion relation be non-linear ?}


\author[W. Bietenholz]{Wolfgang Bietenholz\inst{1,2,}%
  \footnote{Corresponding author\quad E-mail:~\textsf{bietenho@ifh.de},
            Phone: +49\,33\,76277195,
            Fax: +49\,33\,76277330}}
\address[\inst{1}]{Institut f\"{u}r Theoretische Physik,
  Universit\"{a}t Regensburg, D-93040 Regensburg, Germany}
\address[\inst{2}]{Instituto de Ciencias Nucleares, Universidad
  Nacional Aut\'{o}noma de M\'{e}xico, \\ A.P. 70-543, C.P. 04510
  Distrito Federal, Mexico}
\begin{abstract}
  The free photon dispersion relation is a reference quantity for high 
  precision tests of Lorentz Invariance. We first outline theoretical
  approaches to a conceivable Lorentz Invariance Violation (LIV). 
  Next we address phenomenological tests
  based on the propagation of cosmic rays, in particular
  in Gamma Ray Bursts (GRBs). As a specific concept,
  which could imply LIV, we then focus on
  field theory in a non-commutative (NC) space, and we present
  non-perturbative results for the dispersion relation
  of the NC photon.
\end{abstract}
\maketitle                   






\section{Lorentz Invariance Violation (LIV)}

Lorentz Invariance (LI) plays a central r\^{o}le in relativity:
it holds as a global symmetry in Special Relativity Theory,
and as a local symmetry in General Relativity Theory.
We refer to the former in order to be able to describe particle
physics in terms of quantum field theory, where a field $\Phi$ 
(scalar, vector, tensor or spinor) is transformed globally
in some representation $D$ of the Lorentz group $SO(1,3)$, 
\be
\Phi (x) \to U(\Lambda )^{\dagger} \Phi (x) U(\Lambda ) = 
D(\Lambda ) \Phi (\Lambda^{-1} x) \ , \quad 
U ~ {\rm unitary} \ , \quad \Lambda \in SO(1,3) \ .
\ee
If a local field theory obeys LI, then it is also CPT
invariant \cite{CPT}. If LI is broken, CPT may be intact or not.
If, however, CPT is broken, then a LI Violation (LIV) is inevitable
\cite{Greenberg}. Therefore LI can be probed indirectly by testing
CPT predictions, such as the equivalence of the mass and magnetic
moment of particles and their antiparticles. In particular the
masses of $K^{0}$ and $\bar K^{0}$ coincide to a relative precision 
$< 8 \cdot 10^{-19}$ \cite{databook}. 

Specific direct LI tests are even more stringent. In the framework of the
Standard Model Extension (SME) \cite{SME}, local terms are added 
to the Standard Model Lagrangian,
which can cause a spontaneous LIV, {\it e.g.}
\be
{\cal L} = {\rm i} \bar \psi \gamma_{\mu} \partial^{\mu} \psi
- g \bar \psi \phi \psi - {\rm i} \bar \psi (g' G_{\mu \nu} + g''
H_{\mu \nu} \gamma_{5}) \gamma^{\mu} \partial^{\nu} \psi + \dots
\ee
Spontaneous Symmetry Breaking (SSB) of the Higgs field 
$ \phi \in \C^{2}$
leads to the fermion mass.
In analogy the new tensor fields $G$ and $H$ may also undergo SSB, 
which affects LI (unlike the SSB of the Higgs field).
For a simple change of the observers frame, $G$ and $H$ are
transformed as well, so that ${\cal L}$ remains invariant; but 
the fermion $\psi$ perceives them as an invariant background.
LIV is then manifest for instance 
through a modified fermion dispersion relation,
depending on the Yukawa-type couplings $g'$, $g''$.
Kosteleck\'{y} {\it et al.} identified more than 100 parameters
of this kind (for a review, see Ref.\ \cite{Bluhm}), which are 
claimed to be compatible
with the usual properties of the Standard Model except for LI 
(and in part CPT) \cite{ColKosCPT}. This corresponds to a
low energy expansion, after lifting the usual LI condition.
This collaboration interprets 
the photon as a Nambu-Goldstone boson \cite{SSBinSME}.
All LIV parameters emerging in this framework are experimentally
bounded by $O(10^{-27})$ (in a dimensionless form, which factors 
out the Planck scale) \cite{LIVbounds}. If some of these
parameters are non-zero, they still allow for sizable LIV effects
at tremendous energies, though this scenario poses a severe hierarchy 
problem \cite{LIVfinetune}.

In an effective approach, Coleman and Glashow considered
{\em Maximal Attainable Velocities} 
(MAVs, attained in the high energy limit), 
which could slightly differ from the speed of light, 
depending on the particle type 
\cite{ColGla}. They were most interested in the
impact on cosmic rays (which attain the highest energies in 
the Universe). In particular the prediction that proton
rays cannot exceed the Greisen-Zatsepin-Kuz'min (GZK) energy of
$E_{\rm GZK} \approx 6 \cdot 10^{19} ~ {\rm eV}$ over a distance of more 
than $\sim 100 ~ {\rm Mpc}$ in the Cosmic Microwave Background (CMB)
(mainly due to photopion production via a $\Delta$-resonance) 
could be evaded by tiny MAV differences.
However, recent observations favour the validity of the GZK
energy cutoff \cite{GZKholds}: up to $E_{\rm GZK}$ the cosmic ray
flux follows closely a behaviour $\propto E^{-2.7}$, but beyond 
$E_{\rm GZK}$ there is an extra suppression compared to this power
law. This suggests that LI holds even under boosts of Lorentz factors 
like $\gamma = E_{\rm GZK}/ m_{\rm proton} \sim 10^{11}$, {\it i.e.}\
6 orders of magnitude beyond the $\gamma$-factors
accessible in laboratories, which further constrains the MAV differences
\cite{ScuSte08}. 

This issue is reviewed in Ref.\ \cite{cosmo}, covering theoretical
and phenomenological aspects and the current status, 
along with MAV effects in cosmic rays with respect to
decays, \v Cerenkov radiation and neutrino 
oscillation \cite{nuMAV} (these effects were addressed already in Ref.\ 
\cite{ColGla}, and reviewed earlier in Ref.\ \cite{JLM}).

\section{Cosmic $\gamma$-rays}

The highest energies in cosmic $\gamma$-rays have been observed around
$E_{\gamma} \sim 50 ~ {\rm TeV}$. They were emitted from the {\em Crab 
Nebula} \cite{crab} (a remnant of a supernova of the year 1054, at a 
distance of $2~{\rm kpc}$). With certain assumptions about the
origin of these rays (synchrotron radiation of high energy electrons 
and positrons, followed by powerful inverse
Compton scattering) one obtains stringent LIV constraints \cite{JLM}.

The most energetic $\gamma$-rays reaching us from outside our 
galaxy originate from {\em blazars}, a subset of the Active Galactic Nuclei. 
Two prominent blazars are denoted as {\em Markarian 501 and 421} 
at distances of about $157~{\rm Mpc}$ and $110~{\rm Mpc}$, which 
had emitted photons that arrived at Earth
with $E_{\gamma} \sim 20~{\rm TeV}$ \cite{HEGRA} resp.\ 
$E_{\gamma} \sim 10~{\rm TeV}$ \cite{Mkn421}.
These multi-TeV $\gamma$-rays have been considered a puzzle 
similar to the GZK cutoff (which seemed to disagree with the data
in the 20$^{\rm th}$ century): they give rise to the creation of 
electron/positron pairs in the CMB , which can perform inverse
Compton scattering on further CMB photons. This leads to a cascade of
high energy photons, which are attenuated in this way. The penetration
depth of these cascades in the CMB, above certain energy thresholds,
is reviewed for instance
in Ref.\ \cite{BhaSig}: it is sensitive to details of the extragalactic
magnetic fields and the radio background, which are not known precisely.
In analogy to the GZK energy
cutoff, also here LIV effects --- and in particular MAV differences ---
have been proposed as an explanation for the
possibly puzzling $\gamma$-radiation from blazars \cite{GammaTeV}. 
However, also in this case it is not clear if they really represent 
a puzzle ---  {\it e.g.}\ the analysis in Refs.\ \cite{steck} 
suggest that their understanding does not require new physics. 

Since 1967 we also know about the existence of {\em Gamma Ray Bursts}
(GRBs), which are emitted in powerful eruptions. Temporarily they
yield the brightest spots in the sky. Typically they last for 
a few seconds or minutes, and they involve photons in the energy range
$E_{\gamma} \approx (10^{4} \dots 10^{8}) ~ {\rm eV}$. About their origin
we only have speculations, such as the merger of black holes or
neutron stars, along with fireball shock models for the GRB generation.
In any case their existence is suggestive for a high precision test
of the {\em photon dispersion relation,} which represents a purely 
kinematic facet of LI \cite{Camel}. 
A deviation from the linear dispersion relation, {\it i.e.}\
a photon group velocity $v_{\gamma} (E_{\gamma}) \neq const.\,$, can
be obtained most simply by introducing a photon mass $m_{\gamma} > 0$ 
(through a modified Higgs mechanism, keeping the photon MAV unaltered). 
However, the bound based on laboratory experiments, 
$m_{\gamma} < 10^{-18}~{\rm eV}$ \cite{databook}, is so 
tiny that its impact would still not be detectable in GRBs \cite{Camel}.
Hence this approach is not attractive in this context and we return
to $m_{\gamma} = 0$.

\begin{figure}[h!]
\includegraphics[width=\linewidth, height=5cm]{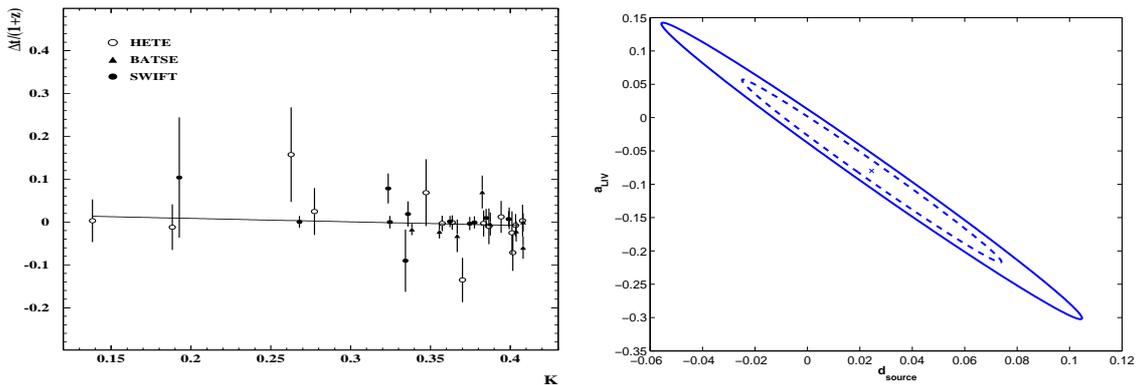}
\caption{Evaluation of the parameters in eq.\ (\ref{dtdEeq}) from
GRB data of three satellites.
The plot on the left shows the left-hand-side of eq.\ (\ref{dtdEeq}) 
against the function $K(z)$. The plot on the right illustrates the
resulting bounds for the parameters $d_{\rm source}$ (relative delay
at the source) and $a_{\rm LIV}$ (LIV effects) with 68 \% resp.\
95 \% C.L. (plots adapted from Ref.\ \cite{Ellis}).}
\label{Ellisfig}
\end{figure}

Instead we switch to the simple effective ansatz (with the 
photon speed $c=1$ attained at low $E_{\gamma}$)
\be
v_{\gamma} \simeq 1 - \frac{E_{\gamma}}{M} \ , 
\ee
where $M$ is a very heavy mass parameter (which could emerge from
some unspecified ``quantum gravity foam''), and we consider only
$O(E_{\gamma}/M)$. Ref.\ \cite{Ellis} confronted this ansatz
with the data of 35 GRBs measured from three satellites,
and observed that photons with higher energy tend to arrive
earlier. They parametrised the observed time delay $\Delta t$
between energy channels, which differ by $\Delta E$, as
\be  \label{dtdEeq}
\frac{ \Delta t}{1 + z} = d_{\rm source} + a_{\rm LIV} \, K(z) \ , \quad
a_{\rm LIV} := \frac{\Delta E}{H_{0} M} \ , \quad
H_{0} ~:~{\rm Hubble~parameter}, ~ z~:~{\rm redshift} \ . \quad
\ee
The parameter $d_{\rm source}$ captures a possible relative delay
already at the source, and $K(z)$ is a function of the redshift,
which is derived in Ref.\ \cite{Ellis}. 
The question is now whether or not the data 
are compatible with a vanishing LIV parameter.
Fig.\ \ref{Ellisfig} (on the left) shows that they do certainly
not rule out $a_{\rm LIV}=0$, although the best linear fit has
a slight slope. The plot on the right specifies the windows for
the parameters $d_{\rm source}$ and $a_{\rm LIV}$ with 68 \% resp.\
95 \% confidence level (C.L.). 
The observation that $a_{\rm LIV}$ cannot deviate much from zero
implies \cite{Ellis}
\be
|M| > 1.4 \cdot 10^{25} ~ {\rm eV} \approx 0.001 \ M_{\rm Planck}
\qquad {\rm (with ~ 95 ~ \% ~ C.L.) \ .}
\ee
This result appears robust due to the large data set considered.
Similar studies based on single GRBs or {\em blazar flares} (again 
involving Markarian 501) did not report evidence for LIV either, and
derived bounds for $|M|$ even up to $O(0.01) \cdot M_{\rm Planck}$
\cite{BWHC}. 

As another purely kinematic effect --- which does not require any
assumptions about the interactions in a LIV modified form of QED ---
there could be {\em vacuum birefringence,} if $v_{\gamma}$ is slightly
helicity dependent,
\be
v_{\gamma}^{\pm}(k) = 1 \pm \frac{k}{2 M_{\rm h}} \qquad
(k = | \vec k \, | ~:~{\rm momentum}, ~ M_{\rm h}~:~{\rm heavy~mass}, ~ 
\pm~:~ {\rm helicities}) \ .
\ee
In this case linearly polarised light should be depolarised
completely after a long enough time of flight.
However, cosmic $\gamma$-rays from quasars at a distance 
of $300~{\rm Mpc}$, with wave lengths of $(400 \dots 800)~{\rm nm}$ 
were observed, essentially with a linear polarisation (less than
$10^{\circ}$ deviation from the polarisation plane) \cite{polar}.
This implies for that LIV scenario even a strongly trans-Planckian 
bound: $|M_{\rm h}| > 10^{4} \, M_{\rm Planck}$ \cite{bireastro}.

Still better precision is expected soon from GRB and blazar flare
photons with $E_{\gamma} \approx 8~{\rm keV} \dots 300~{\rm GeV}$,
to be detected by the Fermi Gamma-ray Space Telescope
\cite{Lamon}, which arrived in orbit in June 2008. 
In addition the Chinese-French Space-based multi-band astronomical 
Variable Objects Monitor mission (SVOM) \cite{SVOM} plans to 
monitor about 80 GRBs a year, starting in 2013. 

Also the subjects of this section are reviewed more extensively in
Refs.\ \cite{JLM,cosmo}.

\section{$U(1)$ gauge theory in a non-commutative (NC) plane}

We now address another theoretical concept for LIV, which is
fundamentally different from the low energy effective approaches
\cite{SME,ColGla} that we referred to in Section 1.
We first consider a Euclidean plane
given by Hermitian coordinate operators
$\hat x_{1}, \, \hat x_{2}$, which do not commute.
This yields a purely spatial uncertainty relation,
\be  \label{NCeq}
[ \hat x_{\mu} , \hat x_{\nu} ] 
= {\rm i} \theta \epsilon_{\mu \nu} \ ,
\quad \Delta x_{1} \cdot \Delta x_{2} \gsim \frac{1}{2} | \theta | \ .
\ee
Here we assume $\theta = const.$ (which can be generalised, of course).
$\sqrt{|\theta |}$ now represents a kind of ``minimal length'',
as a new constant of Nature, like
the parameters $1/M$ or $1/M_{\rm h}$ in Section 2. It might have
an interpretation as the event horizon of a mini black hole 
in a (hypothetical) attempt to measure distances of 
$O(L_{\rm Planck})$ (which would require a tremendous 
energy density) \cite{DFR}.
Field theory on such a space is necessarily {\em non-local} over this
range. This gives rise to the mixing between UV and IR divergences,
which makes perturbation theory mysterious (for a review, see {\it e.g.}\ 
Ref.\ \cite{Szabo} and references therein).

Here we refer to a fully {\em non-perturbative} approach, which starts by
introducing a {\em lattice structure.} Let the spectra of $\hat x_{1}$
and $\hat x_{2}$ be discrete multiples of some ``lattice spacing''
$a$ (although there are no sharp lattice sites, since $\hat x_{1}$,
$\hat x_{2}$  cannot be diagonalised simultaneously). On the other 
hand, the momentum components $k_{1}, \, k_{2}$ commute, and they obey 
the usual periodicity over the Brillouin zone,
\be
e^{{\rm i} k_{\mu} \hat x_{\mu}} = 
e^{{\rm i} (k_{\mu} + 2\pi/a) \hat x_{\mu}} 
\quad \Rightarrow \quad
\hat 1 \!\! 1 =e^{{\rm i} (k_{\mu} + 2\pi/a) \hat x_{\mu}} \
e^{-{\rm i} k_{\nu} \hat x_{\nu}} = \hat 1 \!\! 1 \
e^{ {\rm i} \pi \theta (k_{2}-k_{1})/a} 
\quad \Rightarrow \quad
\frac{\theta}{2a} k_{\mu} \in Z \!\!\! Z \ . \nonumber \qquad
\ee
For fixed parameters $\theta$ and $a$ the momenta turn out to
be discrete, {\it i.e.}\ the lattice must be periodic.
Let us assume periodicity over a $N \times N$ lattice, hence the
$k_{\mu}$ are integer multiples of $2\pi /(aN)$, and from the
computation above we infer
\be
\theta = \frac{1}{\pi} N a^{2} \ .
\ee
In our studies we were therefore interested in the
{\em Double Scaling Limit} (DSL), which takes $a \to 0$ {\em and} 
$N \to \infty$, while keeping $Na^{2} = const.$
This simultaneous UV and IR limit leads to a continuous NC plane of
infinite extent. The necessity to link these limits is related
to the {\em UV/IR mixing} mentioned above.
In other regularisation schemes it is less
obvious how to achieve this delicate balance.

It is equivalent to return to coordinates $x_{\mu}$,
but perform all field multiplications by a star product, {\it e.g.}
\be  \label{starproduct}
\phi (x) \star \psi (x) := \phi (x) \exp \Big( \frac{\rm i}{2} 
\backderi \ \!\!\! _{\mu} \, \theta \, \epsilon_{\mu \nu} \! 
\forderi \ \!\!\! _{\nu} \Big) \psi (x) \ .
\ee
Thus the NC $U(1)$ pure gauge action takes the star-gauge invariant form
\be  \label{2dact}
S [A] = \frac{1}{2} \int d^{2}x \, F_{\mu \nu} \star F_{\mu \nu} \ ,
\quad F_{\mu \nu} = \partial_{\mu} A_{\nu} - \partial_{\nu} A_{\mu}
+ {\rm i} g (A_{\mu} \star A_{\nu} - A_{\nu} \star A_{\mu}) \ ,
\ee
which involves a Yang-Mills term. Its lattice discretisation
does not appear simulation-friendly; note that compact
link variables would have to be star-unitary.
However, there is an exact map onto a matrix model in one point ---
the {\em twisted Eguchi-Kawai model} \cite{GAO} --- which is
Morita equivalent \cite{Morita}. This renders simulations
feasible, and it also provides a sound definition of Wilson loops. 
These Wilson loops are complex in general, but
the lattice action is real positive, since it sums over both 
plaquette orientations.

Our Monte Carlo simulations --- using a heat-bath algorithm --- revealed 
the following features \cite{2dNCU1}: at {\em small area} the Wilson loops 
$\langle W \rangle$ 
of all shapes are real and they follow an area law behaviour.
At {\em large area,} however, a complex phase sets in, which 
obeys for many shapes the simple relation 
${\rm arg} \langle W \rangle = {\rm area} / \theta$.
If we formally identify $1/ \theta$ with a magnetic field
across the plane, then this is the behaviour of the Aharonov-Bohm
effect. This identification can be conjectured on tree level,
and it has been used extensively in the literature; for instance,
it is inherent in the Seiberg-Witten map 
\cite{SeiWit}. In the case discussed above, however, this relation 
only shows up as a non-perturbative effect.

Moreover the Wilson loops of large area 
were observed to be {\em shape dependent,} so the symmetry under
Area Preserving Diffeomorphisms is broken, including the
subgroup of linear unimodular transformations, $SL(2,R)$ 
\cite{BBT}. That symmetry was a corner stone 
in the solution of Yang-Mills theories in a commutative plane.
We conclude that the same technique cannot be applied in the
NC plane, so an analytical solution would be difficult,
but the system may have a rich structure.
\vspace*{-2mm}

\section{The fate of the non-commutative photon}

It is straightforward to extend action (\ref{2dact})
to higher dimensions and any $U(n)$ gauge group in the 
fundamental representation, but the
formulation of $SU(N)$ gauge theories in NC spaces fails.
Therefore 4d NC $U(1)$ gauge theory is appropriate to 
investigate a link to particle phenomenology.

On the perturbative side, the 1-loop calculation of the effective
potential suggests a photon dispersion relation of the form \cite{MST}
\vspace*{-1mm}
\be  \label{disp}
E^{2} = \vec p^{\, 2} + C \, \frac{g^{2}}{p \circ p } \ , \quad 
p \circ p := - \Theta^{\mu \nu} p_{\nu} \Theta_{\mu}^{\ \ \sigma} 
p_{\sigma} \geq 0 \ , \quad ( \ C = const.
, \ \ \Theta^{\mu \nu} = - {\rm i} \, [\hat x^{\mu}, \hat x^{\nu}] 
\ ) \ , \quad
\ee
which contains an {\em IR divergence} at finite 
$\Theta$. Moreover we see that the transition to a vanishing
non-commutativity tensor, $\Theta \to 0$, is {\em not
smooth} (which invalidates an expansion in small $\Vert \Theta \Vert$).  
Therefore one could even feel tempted to claim a {\em lower} bound 
for $\Vert \Theta \Vert$ --- {\em if} it is non-zero ---
from the fact that {\em no} deviation from the linear dispersion has
been observed down to small photon momenta \cite{CamelNC}.

However, the situation becomes even more dramatic due to the
observation that the above 1-loop correction is actually
{\em negative}, $C = - 2/\pi^{2}$ \cite{LLT}.\footnote{On the other
hand, a positive IR divergence has been observed numerically in the 
NC $\lambda \phi^{4}$ model \cite{NCphi4}.}
This looks worrisome indeed: the theory seems to be IR unstable.
Without a ground state the system does hardly have a 
physical interpretation. The negative IR divergence is
avoided in a supersymmetric version, but here we are interested
in the NC photon without a SUSY partner. We addressed the issue 
of its stability in a fully non-perturbative study \cite{NCQED}.

To this end we considered a Euclidean space consisting of a commutative
and a NC plane. The former contains the Euclidean time. This is
essential in order to preserve reflection positivity as one of
the Osterwalder-Schrader axioms, which are required for the
transition to Minkowski signature. Once time is commutative,
non-commutativity can be restricted to one plane by means of 
rotations. 
This setting is the safest NC formulation beyond perturbation theory,
and it can be viewed as minimal non-commutativity; there is again
just one non-commutativity parameter $\theta$, as in eq.\ (\ref{NCeq}).

\begin{figure}[h!]
\begin{center}
\includegraphics[angle=270,width=.34\linewidth]{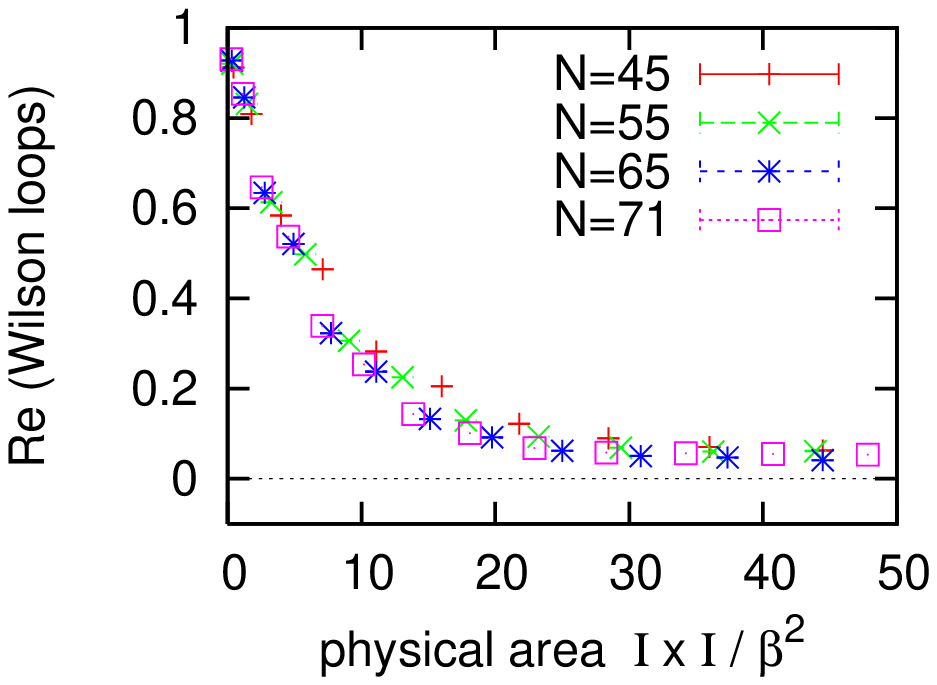} \hspace*{-5mm}
\includegraphics[angle=270,width=.34\linewidth]{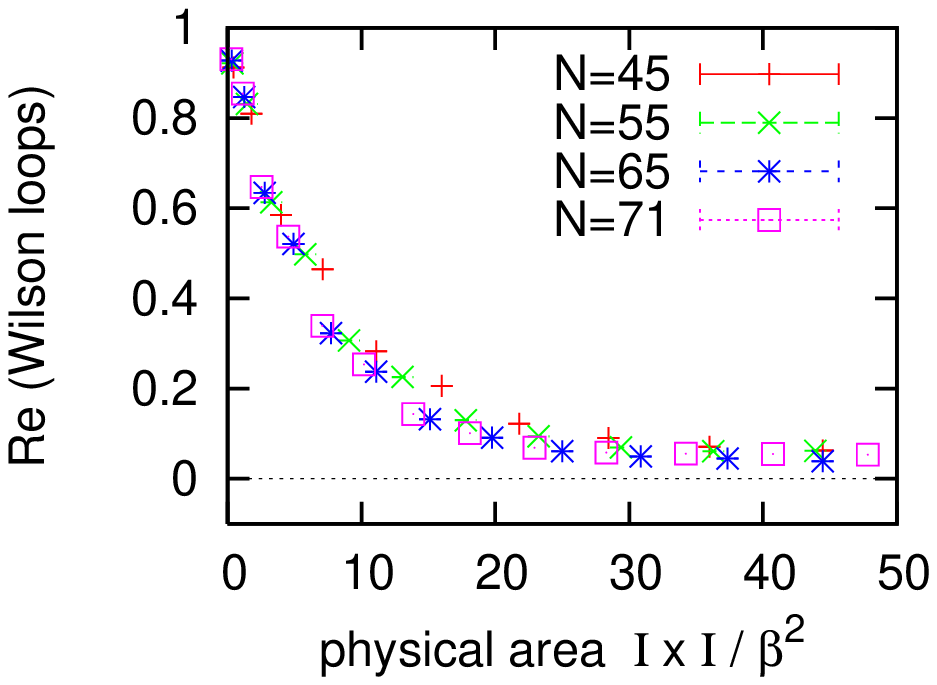} \hspace*{-5mm}
\includegraphics[angle=270,width=.34\linewidth]{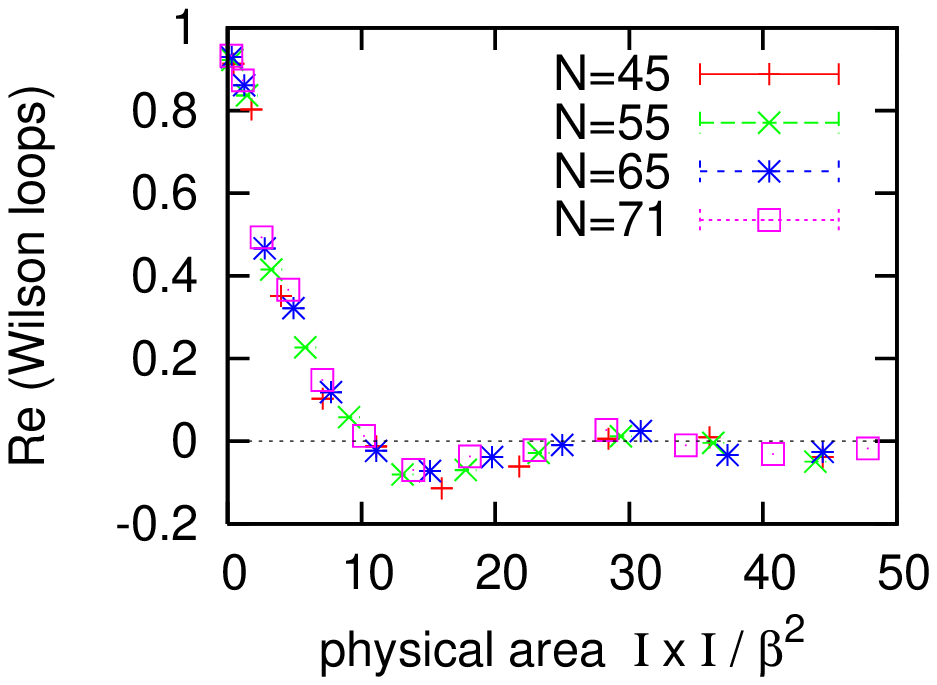}
\vspace*{-3mm} \\
\end{center}
\caption{{\it Wilson loops measured in NC QED$_{4}$ at 
$N /\beta^{2} = 20$ in the commutative plane, a mixed plane and
the NC plane (from left to right). In the first two planes
$\langle W \rangle$ is real, whereas the oscillation of 
$\, {\rm Re} \, \langle W \rangle \,$ in the NC plane is due to a 
rotating complex phase (similar to NC QED$_{2}$ \cite{2dNCU1}).}} 
\label{4dWl}
\end{figure}

We regularise the commutative plane by a standard $L \times L$
lattice, and the NC plane with an $N \times N$ lattice ($L \approx N$),
which is again mapped onto $N \times N$ matrices in the twisted 
Eguchi-Kawai model. The first goal in our numerical study \cite{NCQED}
was the identification of a DSL. The results for a set of observables
(Wilson loops and lines and their 2-point functions, matrix spectra
as coordinates of a dynamically generated geometry) stabilised
to a good approximation if different $N$ and 
$\beta \propto 1/ g^{2}$ were tuned as
\be  \label{4dDSL}
N / \beta^{2} = const. \qquad \Rightarrow \qquad
a \propto 1 / \beta \qquad {\rm and} \qquad
\theta \propto N / \beta^{2} \ .
\ee
As an example we show in Fig.\ \ref{4dWl} the real part of
square shaped Wilson loops in different planes, at a fixed
ratio $N / \beta^{2} =20$. In the commutative plane and in the
mixed planes $\langle W \rangle$ is real due to reflection
symmetry on the commutative axes.
In the NC plane we see an oscillating behaviour of the real part
at large area, which is due to a rotating complex phase, similar 
to the 2d case \cite{2dNCU1} that we described in Section 3.

\begin{figure}
\vspace*{-4mm}
\begin{center}
\includegraphics[angle=270,width=.45\linewidth]{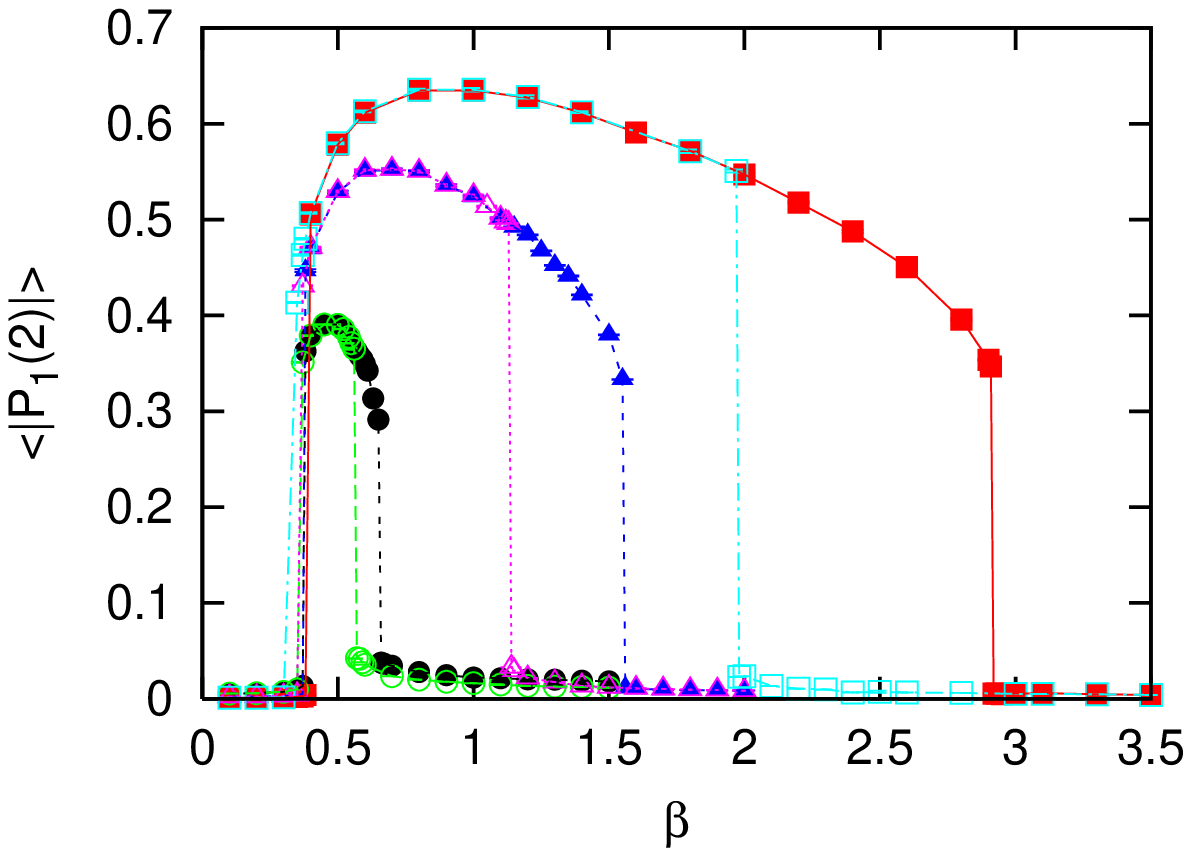}
\includegraphics[angle=270,width=.45\linewidth]{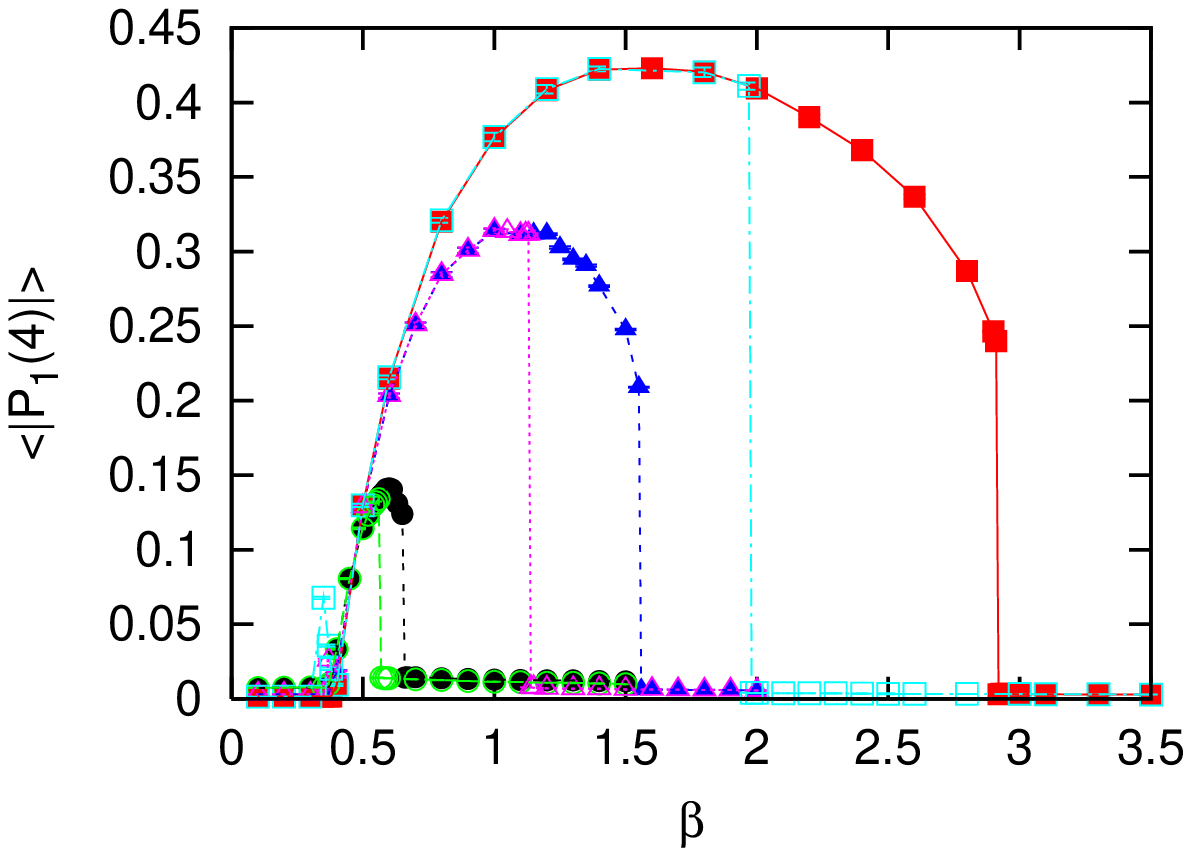}
\vspace*{-3mm} \\
\end{center}
\caption{{\it The expectation values of open Wilson lines 
of length $2$ and $4$ lattice spacings (on the left/right) for
$N=15, \ 25 $ and $35$ (curves connecting bullets, triangles
and squares, with hysteresis
at the transition to the weak coupling phase). At intermediate
gauge coupling we discovered a phase of spontaneously
broken translation symmetry.}}
\label{Polyfig}
\vspace*{-3mm}
\end{figure}

\begin{figure}
\begin{center}
\includegraphics[angle=0,width=.58\linewidth]{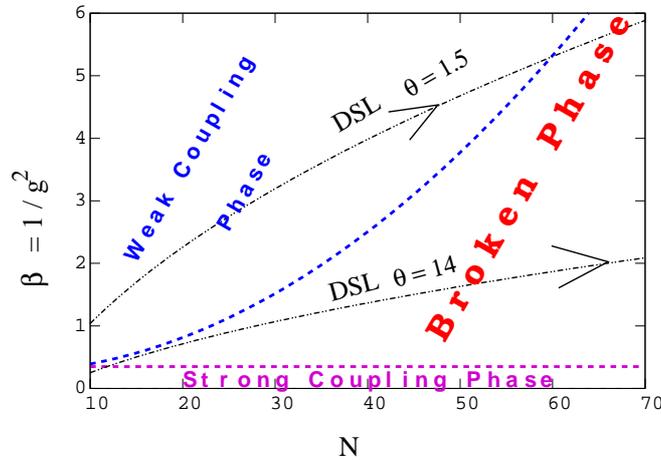}
\vspace*{-5mm} \\
\end{center}
\caption{{\it The phase diagram for QED$_{4}$ with a NC plane
(in lattice units):
between the strong coupling phase ($\beta \lsim 0.35$) and the weak
coupling phase we recognise a phase of broken translation invariance.
The weak/broken transition rises like $\beta \propto N^{2}$, whereas
the DSL for a fixed NC parameter $\theta$ follows $\beta \propto
\sqrt{N}$, so any DSL ends up in the broken phase.}} 
\vspace*{-1mm}
\label{phasedia}
\end{figure}

The next goal was the exploration of the {\em phase diagram.}
For this purpose we measured expectation values of open Wilson lines 
$P_{\mu}(n)$ in the NC plane, 
with a length $\Theta_{\mu \nu} p_{\nu} = n a \hat \mu$ 
of a few lattice spacings ($| \hat \mu | = 1, \ n \in \! \N$).
The $P_{\mu}(n)$ are gauge invariant 
in NC space; note that star-gauge transformations are non-local 
themselves. Since these open Wilson lines carry finite momentum, 
they are suitable order parameters to detect the spontaneous breaking 
of translation symmetry.
Results for $N=15, \, 25$ and $35$ are shown in Fig.\ \ref{Polyfig}.
Translation symmetry is intact at strong and at weak gauge coupling,
in agreement with the corresponding expansions, but it is broken at
intermediate coupling. This effect is purely non-perturbative; it
was not visible in analytical studies. The transition
strong/moderate coupling occurs at $\beta \approx 0.35$ (for any $N$),
whereas the transition moderate/weak coupling roughly follows a
behaviour $\beta \propto N^{2}$ (that phase transition is
of first order, as we see in Fig.\ \ref{Polyfig} from the hysteresis).
The existence of a broken symmetry phase at moderate coupling
was also confirmed in simulations of the closely related 4d
twisted Eguchi-Kawai model \cite{4dTEK}. In our case we obtain the
phase diagram in Fig.\ \ref{phasedia}.
The DSL is attained by following lines of constant $\theta$.
Relation (\ref{4dDSL}) implies that they correspond to
$\beta \propto \sqrt{N}$, hence the DSL always leads to the 
{\em broken phase.} Since we found stability for a number of
observables as we approach this limit, it seems to provide 
IR stability for the NC photon.

\begin{figure}
\vspace*{-3mm}
\begin{center}
\includegraphics[angle=270,width=.49\linewidth]{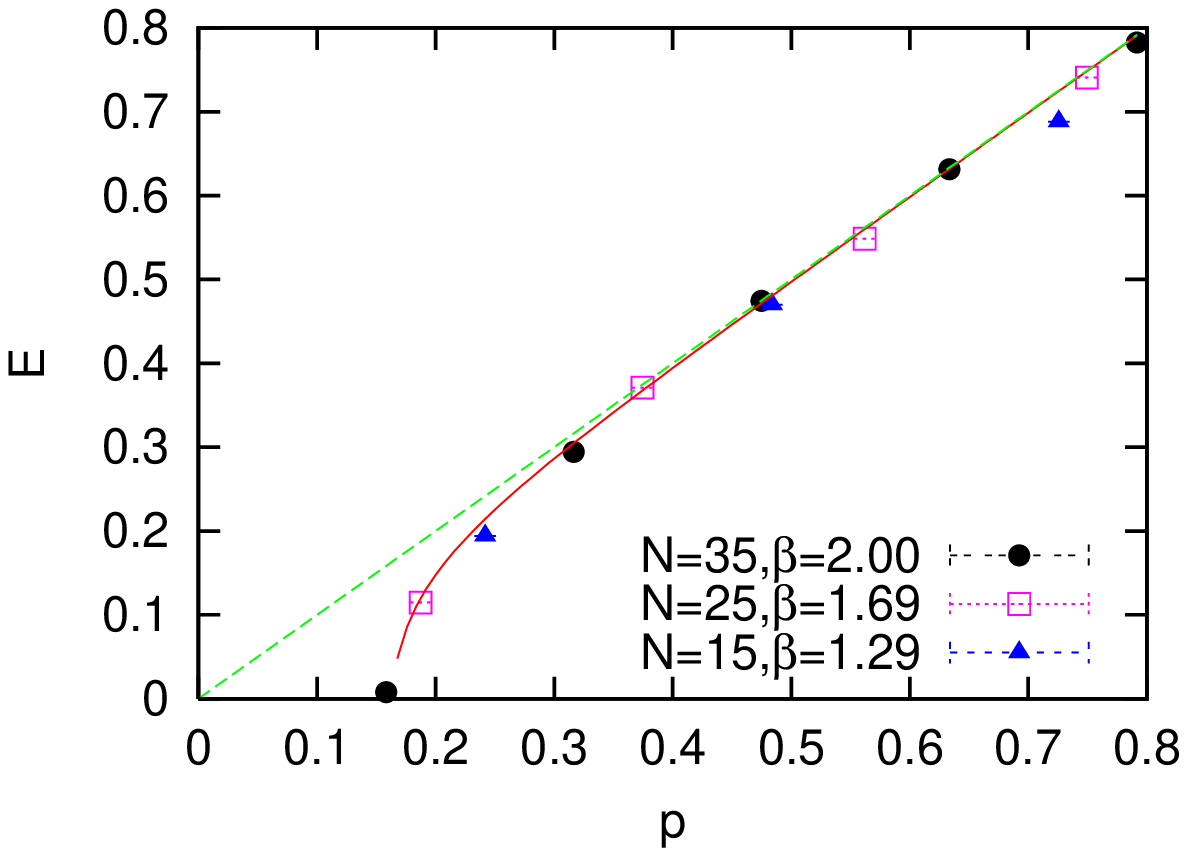}
\includegraphics[angle=270,width=.49\linewidth]{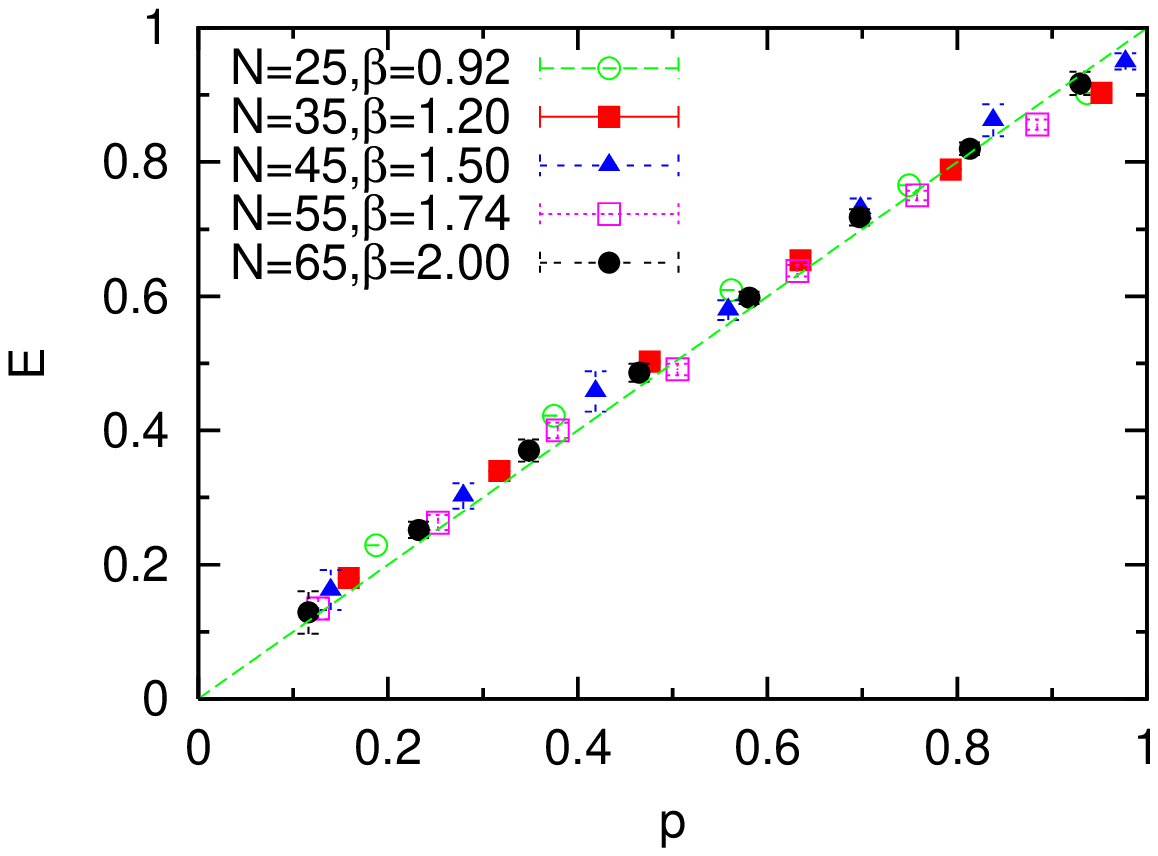}
\vspace*{-2mm} \\
\end{center}
\caption{{\it Dispersion relations $\, E(p) \ $
for the photon in a NC space, in the weak coupling phase (on the left), 
and in the broken phase (on the right).
The former result is consistent with the perturbatively predicted IR
instability (``tachyonic'' behaviour). In the physical phase, however,
the photon is massless again, corresponding to a Nambu-Goldstone boson 
of the broken translation symmetry.}}
\label{disprel}
\vspace*{-5mm}
\end{figure}

At last we consider non-perturbative results for the {\em NC photon 
dispersion relation.} The plot in Fig.\ \ref{disprel} on the left shows the
behaviour in the symmetric phase at weak coupling: it is fully consistent
with the perturbative result of a negative IR divergence. However, we saw 
that the DSL leads to the broken phase, which is therefore physically 
relevant. For that phase the plot on the right of Fig.\ \ref{disprel}
reveals an {\em IR stable} behaviour, at least for the energy depending
on the momentum component in the commutative space direction
only, $E (p=p_{3})_{p_{1}=p_{2}=0}$.
Therefore the photon could survive the transition to an NC world
after all. Then one could again try to interpret it as the
Nambu-Goldstone boson of the SSB of Poincar\'{e} symmetry, similar to
the interpretation in the context of the SME \cite{SSBinSME} (cf.\ Section 1).
\vspace*{-3mm}

\section{Conclusions}

Cosmic $\gamma$-rays enable LI tests to extreme precision.
So far LIV has not been detected anywhere --- we reviewed
failed attempts based on GRB data.

As theoretical frameworks leading to LIV we discussed on one side
low energy effective theories \cite{SME,Bluhm,ColKosCPT,SSBinSME}
and an effective ansatz which introduces
particle specific Maximal Attainable Velocities \cite{ColGla}.
Another approach is based on NC geometry --- the presence of
IR divergences in quantum field theory on such spaces clarifies
its distinction from low energy effective theories.

In Section 4 we were concerned with the kinematics of a NC photon.
Perturbation theory to one loop suggests its IR instability 
\cite{LLT}, which looks like a disaster for NC field theory.
The picture changes, however, in light of a non-perturbative
consideration.

Our numerical study revealed a phase of spontaneously broken
Poincar\'{e} symmetry, which is not visible
to perturbative calculations. In fact the limit to the
continuum and to infinite volume at fixed non-commutativity
(the Double Scaling Limit) leads to this phase of broken symmetry,and 
{\em not} to the symmetric weak coupling phase. In the broken symmetry
phase we observed stability for a number of observables \cite{NCQED}.
In particular the photon dispersion relation in a commutative
sub-space appears linear. However, its behaviour at 
non-zero momentum components in the NC directions 
is not explored yet. In that respect, non-perturbative results 
could be confronted with phenomenological data --- in particular
from GRBs --- to establish a robust bound on $\Vert \Theta \Vert$
in Nature. 
On the phenomenological side, more stringent data are expected
from new large-scale GRB observations, for instance in the
projects discussed in Refs.\ \cite{Lamon,SVOM}.

Our non-perturbative results so far imply that the photon may survive 
in a NC world --- despite the IR disaster on the 1-loop level ---
without the requirement to proceed to SUSY.
 
\begin{acknowledgement}

It was my pleasure to attend the 4$^{\rm th}$ EU RTN 
Workshop on ``Constituents, Fundamental Forces 
and Symmetries of the Universe'' in the beautiful city of 
Varna (Bulgaria), where this talk was presented.
I would like to thank my collaborators for their contributions
to Refs.\ \cite{BBT,NCphi4,NCQED,2dNCU1}, the authors of
Ref.\ \cite{Ellis} for their permission to reproduce Fig.\
\ref{Ellisfig}, and F.W.\ Stecker for a helpful comment.
This work was supported by the Deutsche Forschungsgemeinschaft (DFG) 
through Sonderforschungsbereich SFB/TR55 
``Hadron Physics from Lattice QCD''.

\end{acknowledgement}

\end{document}